\documentclass[10pt,conference]{IEEEtran}
\IEEEoverridecommandlockouts
\usepackage{cite}
\usepackage{amsmath,amssymb,amsfonts}
\usepackage{algorithmic}
\usepackage{graphicx}
\usepackage{textcomp}
\usepackage{xcolor}
\usepackage{makecell}
\usepackage{tcolorbox}
\usepackage{multirow}
\usepackage{url}
\usepackage{arydshln}
\usepackage[numbers,sort&compress]{natbib}

\def\BibTeX{{\rm B\kern-.05em{\sc i\kern-.025em b}\kern-.08em
    T\kern-.1667em\lower.7ex\hbox{E}\kern-.125emX}}
\begin{document}

\newcommand{\ie}{{\textit{i.e.}}, \xspace}
\newcommand{\eg}{{\textit{e.g.}}, \xspace}
\newcommand{\etc}{\textit{etc.} }
\newcommand{\etal}{{\textit{et al.}}}

\newcommand{\finding}[2]{
\begin{tcolorbox}[width=\linewidth,boxrule=0pt,top=1pt, bottom=1pt, left=1pt,right=1pt, colback=gray!20,colframe=gray!20]
\textbf{Finding #1:} 
{#2}
\end{tcolorbox}}

\title{One Adapter for All Programming Languages? Adapter Tuning for Code Search and Summarization\\
}

\author{\IEEEauthorblockN{Deze Wang\IEEEauthorrefmark{2},
Boxing Chen,
Shanshan Li\thanks{\IEEEauthorrefmark{1} Shanshan Li is the corresponding author.}\IEEEauthorrefmark{2}\IEEEauthorrefmark{1},
Wei Luo,
Shaoliang Peng\IEEEauthorrefmark{4},
Wei Dong\IEEEauthorrefmark{2},
Xiangke Liao\IEEEauthorrefmark{2}
}

\IEEEauthorblockA{\IEEEauthorrefmark{2}National University of Defense Technology, Changsha, China
}
\IEEEauthorblockA{\IEEEauthorrefmark{4}Hunan University, Changsha, China\\
\{wangdeze14,shanshanli,wdong,xkliao\}@nudt.edu.cn, chenboxing@gmail.com,  luowei828@163.com, slpeng@hnu.edu.cn
}
}

\maketitle

\begin{abstract}
As pre-trained models automate many code intelligence tasks, a widely used paradigm is to fine-tune a model on the task dataset for each programming language. A recent study reported that multilingual fine-tuning benefits a range of tasks and models. However, we find that multilingual fine-tuning leads to performance degradation on recent models UniXcoder and CodeT5. 

To alleviate the potentially catastrophic forgetting issue in multilingual models, we fix all pre-trained model parameters, insert the parameter-efficient structure adapter, and fine-tune it. Updating only 0.6\% of the overall parameters compared to full-model fine-tuning for each programming language, adapter tuning yields consistent improvements on code search and summarization tasks, achieving state-of-the-art results. In addition, we experimentally show its effectiveness in cross-lingual and low-resource scenarios. Multilingual fine-tuning with 200 samples per programming language approaches the results fine-tuned with the entire dataset on code summarization. Our experiments on three probing tasks show that adapter tuning significantly outperforms full-model fine-tuning and effectively overcomes catastrophic forgetting.

\end{abstract}


\section{Introduction}
Deep learning models are widely applied to tasks of different programming languages, such as code search and summarization in  programming  languages like Java, Python, Ruby, \etc The existing paradigm to solve these tasks is first loading pre-trained language models and then training and evaluating them on language-specific datasets. For a given task, it is necessary to train and maintain separate models for each language.

N  programming languages require N separate models, which are challenging to train, deploy and maintain individually. In addition, researchers in the Natural Language Processing (NLP) field have reported that exposing models to multiple languages improves performance in low-resource languages~\cite{Ha2016TowardMN,Firat2016ZeroResourceTW,Dabre2020ACS}. Unlike natural languages, multiple programming languages have similar grammatical forms, and different monolingual models fine-tuned from the same pre-trained model would share a common vocabulary. Therefore, multilingual training may be more beneficial for knowledge transfer between programming languages than natural languages.

The most related work by Ahmed and Devanbu~\cite{Ahmed2022MultilingualTF} investigates that based on pre-trained code models CodeBERT~\cite{Feng2020CodeBERTAP} and GraphCodeBERT~\cite{Guo2021GraphCodeBERTPC}, multilingual fine-tuning leads to almost consistent improvement for all six programming languages~\cite{Lu2021CodeXGLUEAM} on code search and summarization tasks. However, we experimentally find that this conclusion cannot be generalized to the latest pre-trained code models UniXcoder~\cite{Guo2022UniXcoderUC} and CodeT5~\cite{Wang2021CodeT5IU}. On these models, multilingual fine-tuning leads to performance degradation in most programming languages for code search and summarization tasks. It is challenging to train a model that supports multiple programming languages simultaneously and maintains comparable performance to a dedicated monolingual model.

Many studies have shown that multilingual models may suffer from negative transfer due to catastrophic forgetting of knowledge gained from pre-training~\cite{French1999CatastrophicFI}. We simply fix all parameters of the pre-trained model, insert the adapter~\cite{Houlsby2019ParameterEfficientTL} into the model and fine-tune it. As a result, we find that it can alleviate the above issue. Furthermore, we demonstrate why adapter tuning is effective through probing~\cite{Hewitt2019ASP} analysis experiments. 

We introduce the adapter module to different pre-trained code models. Compared to training the entire model for each programming language, adapter tuning only tunes no more than 0.6\% of the whole parameters. The newly obtained models outperform models fine-tuned with all parameters and achieve state-of-the-art results on code search and summarization tasks. We experimentally find that on the code summarization task, adapter tuning with a small number of samples per language can approach the results of fine-tuning with the entire dataset, demonstrating that it is possible to adapt to a new programming language quickly. Furthermore, we show its effectiveness in cross-lingual tasks and conduct linguistic probing experiments to show why it works. 

The main contributions of our paper are as follows:
\begin{itemize}
\item We experimentally find that multilingual fine-tuning, which obtains significant performance gains on CodeBERT and GraphCodeBERT, leads to performance degradation on UniXcoder and CodeT5.

\item We show evidence that adapter tuning can significantly overcome catastrophic forgetting in multilingual fine-tuning through three probing tasks. 

\item Based on the pre-trained UniXcoder and CodeT5, our models tune no more than 0.6\% of the whole parameters on code search and summarization tasks for six programming languages, obtain consistent performance gains and achieve state-of-the-art results.

\item We show that adapter tuning with 200 samples per programming language ($ \sim $0.02\% of the original dataset) can perform well in the code summarization task.

\end{itemize}

\section{Preliminaries}

\subsection{Fine-tuning}
With the great success of pre-trained models, the pretrain-then-finetune paradigm has become the dominant paradigm in the NLP field. Fine-tuning uses the parameters of the pre-trained model as initialization and quickly adapts to new tasks without training from scratch. It trains the model in a supervised manner. Formally, given a set of task-specific samples $X$ and corresponding labels $Y$, fine-tuning is to find a set of parameters satisfying $\theta = \underset{\theta}{arg~min}~P(Y|X;\theta)$.

In this paper, multilingual training we study here refers to the fine-tuning stage of pre-trained models. It is performed on the combined dataset of all programming languages. The pre-trained models have been trained using multilingual data in an unsupervised manner. Unlike fine-tuning a model using a monolingual dataset and building a model for each language, we expect to fine-tune a multilingual model that is competent for multiple  programming  languages simultaneously. 

\begin{figure}[t]                         
	\centering                              
	\includegraphics[width=\linewidth]{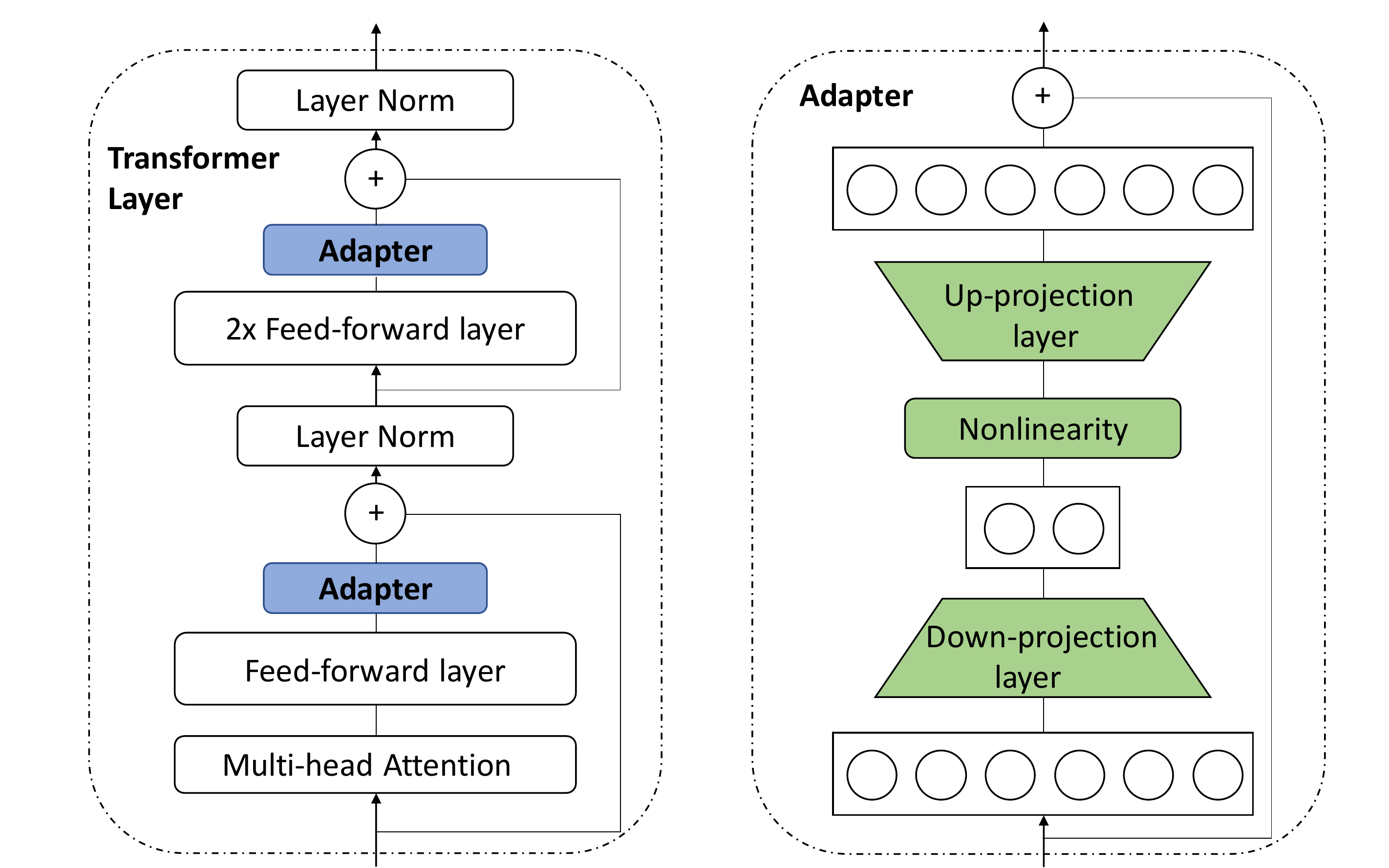} 
	\caption{Architecture of the adapter module and its integration with the Transformer. \textbf{Left}: Adapter is inserted after the multi-head attention and the two feed-forward layers. \textbf{Right}: The adapter contains a bottleneck, which has fewer parameters than the attention and feed-forward layers, and it also contains a skip-connection. The green part of the figure shows the parameters to tune.}
	\label{pic_adapter}                              
\end{figure} 

\subsection{Adapters}

When the prevalent paradigm of transfer learning is to fine-tune all parameters of a pre-trained model, the adapter module provides a lightweight alternative to update only a small number of extra parameters while keeping pre-trained parameters frozen.

Fig.~\ref{pic_adapter} illustrates the standard adapter architecture~\cite{Houlsby2019ParameterEfficientTL}. For a Transformer-based architecture~\cite{Vaswani2017AttentionIA}, a small set of new parameters is introduced in every transformer layer. It is inserted behind the attention and feed-forward layer of each transformer layer, respectively. The adapter contains two projection layers and a nonlinear layer. A skip-connection layer is employed across the adapter. Given a hidden input vector $h$, the output of the adapter is:
\begin{equation}
Z = W_{Up}(\sigma(W_{Down}(h)))+ h 
\end{equation}
where $\sigma$ is the activation function, $W_{Up}\in R^{m\times d}$ and $W_{Down}\in R^{d\times m}$ are parameters of projection layers. $d$ is the hidden size of the Transformer model, and $m$ is the dimension of the adapter. As shown in Fig. 1, the dimension of $m$ is generally smaller than $d$. The two projection layers form a bottleneck structure.

\section{Research Method}
To address multilingual tasks for source code understanding and generation, we investigate a series of research questions and describe their study design.

\subsection{RQ1: How Does Multilingual Fine-tuning Perform on Different Models and Downstream Tasks?} Ahmed and Devanbu~\cite{Ahmed2022MultilingualTF} find that multilingual fine-tuning can benefit CodeBERT and GraphCodeBERT, and we investigate whether this finding can be generalized to other pre-trained models.

\textbf{RQ1 Design:} We fine-tune UniXcoder and CodeT5 with the multilingual dataset on code search and code summarization tasks. The multilingual dataset is a combination of datasets in six programming languages~\cite{Lu2021CodeXGLUEAM}, including Python, Java, JavaScript, Ruby, Go, and PHP. On code search and code summarization, we evaluate the performance of models fine-tuned on the multilingual dataset and compare them with models fine-tuned on the monolingual datasets. To check whether multilingual fine-tuning performs consistently across different pre-trained code models, we also compare the results of CodeBERT, GraphCodeBERT, and PLBART with the same settings~\cite{Ahmed2022MultilingualTF}. 

\subsection{RQ2: How Effective is Adapter Tuning in Cross-lingual Scenarios?} A cross-lingual task is to fine-tune pre-trained models with the dataset in one programming language and test on the dataset in another. By inserting the adapter into pre-trained models and then adjusting only this small number of extra parameters, we are interested in whether adapter tuning is effective in solving cross-lingual tasks.

\textbf{RQ2 Design:} We fine-tune UniXcoder and CodeT5 with six monolingual datasets, and for each monolingual model, we test it on all programming languages separately.
For adapter tuning, we tune an adapter for each language and then test these adapters on all programming languages equally.  We further compare the performance of adapter tuning and full-model tuning on each cross-lingual task.

\subsection{RQ3: How Effective is Adapter Tuning Over Multilingual Full-model Fine-tuning on Different Pre-trained Models?} As cross-lingual tasks require knowledge transfer capabilities in pairs of programming languages, multilingual tasks require a model competent in multiple  programming languages simultaneously. We expect to investigate whether adapter tuning effectively solves multilingual tasks.

\textbf{RQ3 Design:} We insert the adapter into UniXcoder and CodeT5 and fine-tune the adapter with the multilingual dataset. The performance of the adapter is evaluated on code search and code summarization tasks. We compare the performance of multilingual models fine-tuned by the adapter with all other models, including corresponding monolingual models and multilingual models with full-model fine-tuning.

\subsection{RQ4: How Effective is Multilingual Fine-tuning in Low-resource Scenarios?}  For a multilingual model, we expect it to learn and generalize to new programming languages quickly. However, there is limited labeled data for models to learn for many programming languages. Therefore, we sample existing datasets and study the performance of multilingual learning in low-resource scenarios.

\textbf{RQ4 Design:} We randomly sample the datasets in each programming languages, and choose 100, 200, 500 and 1000 samples for each language. Then, we insert the adapter into CodeT5 and evaluate the model on the combinations of these data. We vary the random seed, repeat the experiment several times, and average the results to check the effectiveness of multilingual learning.

\subsection{RQ5: Why is Adapter Tuning Better than Full-model Fine-tuning in the Above Scenarios?} Although adapter tuning shows superior performance in the above scenarios, it does not directly provide insights into why adapter tuning can surpass full-model fine-tuning with very few parameters. We use linguistic probing experiments to explore this point.

\textbf{RQ5 Design:} We employ probing experiments to assess the hidden state embeddings of multiple models and measure their ability to capture fundamental characteristics related to code. We adopt three probing tasks of code length prediction, cyclomatic complexity and invalid type detection~\cite{Karmakar2021WhatDP}. These tasks correspond to probing surface-level, syntactic and semantic information of source code, respectively. Precisely, after fine-tuning various models on downstream tasks, we extract pre-trained vector embeddings on probing tasks to check whether the models reflect the understanding of code information. In particular, we verify whether adapter tuning that shows effectiveness in the above scenarios performs consistently in the probing experiments.

\section{Experimental Setup}
This paper adopts three downstream tasks: code search, code summarization, and low-resource code summarization. We also apply three probing tasks to examine models: code length prediction, cyclomatic complexity and invalid type detection. We next describe the details of pre-trained models, tasks, and datasets.

\subsection{Pre-trained Models}
We choose the state-of-the-art model UniXcoder for code search and the state-of-the-art model CodeT5 for code summarization task. UniXcoder can support both source code comprehension and generation tasks. It controls the model behavior through self-attention masks, applying an encoder-only architecture for code search and an encoder-decoder architecture for code summarization. For CodeT5, it has two versions of CodeT5-small (60M) and CodeT5-base (220M). We use the well-performing CodeT5-base for code summarization.

We also compare the results with those on CodeBERT, GraphCodeBERT and PLBART~\cite{Ahmad2021UnifiedPF} models. CodeBERT and GraphCodeBERT apply the same architecture as RoBERTa~\cite{Liu2019RoBERTaAR} with 125M parameters. CodeBERT is pre-trained with bimodal data of source code and natural language. GraphCodeBERT incorporates data flow information into the pre-training process on top of CodeBERT. These two models are encoder-only models for source code understanding. We follow the previous work and add six transformer layers to them as the decoder for source code generation task. PLBART adopts the encoder-decoder architecture and applies denoising objectives to pre-train the model with source code and natural language. It has 140M parameters for both source code understanding and generation tasks.

\subsection{Tasks}
\paragraph{Code Summarization} Code summarization aims to generate a text summary describing the code. The input to the model is a code snippet, and the output is a natural language description of the code functionality.
\paragraph{Code Search} Given a natural language query as the input, code search task is to find the most semantically related code from a collection of candidate programs. Since CodeT5 does not provide any results on code search, we only utilize UniXcoder for this task.
\paragraph{Code Length Prediction (LEN)} The amount of information in codes may vary across lengths. We expect to use an intuitive task to predict the code length and to check whether different models encode such surface information. For simplicity, this task is converted to a classification task that predicts which of the five length intervals the code sequence falls in.

\paragraph{Cyclomatic Complexity (CPX)} The cyclomatic complexity reflects the structure information of the source code. In order to perform code tasks such as code summarization, it is necessary for models to understand the syntactic structure of input code tokens. This task aims to predict the cyclomatic complexity corresponding to code tokens and check to what extent different models encode structure information. Since the number of linearly independent paths through a code snippet determines the complexity of the code, it may be a challenge to predict it based on the token sequence alone.
\paragraph{Invalid Type Detection (TYP)}  Similar to denoising tasks of BART~\cite{Lewis2020BARTDS,Ahmad2021UnifiedPF}, this task is to distinguish semantically valid code snippets from invalid ones. Invalid samples are constructed by randomly tampering with the original data types in code snippets. The purpose of TYP is to check whether different models can identify invalid data types by code context and further verify to what extent these models understand code semantics.

\begin{table}[t]
	\caption{CodeSearchNet dataset}
	\begin{center}
		
		\begin{tabular}{lcccc}
			\hline
			\makecell[c]{Programming \\ language}&Training&Dev&Test&Candidate codes\\
			 
			\hline
			Ruby & 24,927 & 1,400 & 1,261 & 4,360 \\ 
			JavaScript & 58,025 & 3,885 & 3,291 & 13,981 \\
			Java & 164,923 & 5,183 & 10,955 & 40,347 \\
			Go & 167,288 & 7,325 & 8,122 & 28,120 \\
			PHP & 241,241 & 12,982 & 14,014 & 52,660 \\
			Python & 251,820 & 13,914 & 14,918 & 43,827 \\
			\hline
			
		\end{tabular}
		\label{dataset}
	\end{center}
\end{table}

\subsection{Evaluation Datasets}
\paragraph{Code Summarization} We choose the dataset from CodeXGLUE, which incorporates CodeSearchNet~\cite{Husain2019CodeSearchNetCE} and is carefully de-duplicated. Table~\ref{dataset} shows the statistics of the dataset. This dataset contains pairs of code snippets and natural language descriptions for six programming languages, including Python, Java, JavaScript, Ruby, Go, and PHP. From the table, it can be seen that there is a significant difference in the data size for different programming languages. In particular, the datasets of Ruby and JavaScript are much smaller than the datasets of other programming languages.

\paragraph{Code Search}
This dataset we use is adapted from the same CodeSearchNet dataset with additional candidate codes by Guo~\etal~\cite{Guo2021GraphCodeBERTPC}. Except for the extra candidate codes for retrieval, the dataset is the same as we use for code summarization.
\paragraph{Datasets for Probing Tasks} We adopt datasets from Karmakar and Robbes~\cite{Karmakar2021WhatDP} for probing tasks. In detail, The length labels is set to 5 class-bins (0-50, 50-100,~\textit{etc.}) for LEN task. Complexity labels of CPX are obtained with the metrix++ tool, ranging from 0 to 9. Code snippets are divided into two classes for TYP task based on whether it contains invalid types. Datasets for each task contain 10k samples and are class-balanced.
\subsection{Evaluation Metrics}
\paragraph{Code Summarization}
Following the previous work, we use smoothed BLEU-4~\cite{Papineni2002BleuAM} as the evaluation metric. It is a precision-based metric and measures the n-gram geometric precision between the generated text and ground truth text. We also follow the previous work and average the BLEU across  programming languages to report the overall score.
\paragraph{Code Search}
We use Mean Reciprocal Rank (MRR) as the evaluation metric. MRR is the average of the reciprocal rank of results of a set of queries. The reciprocal rank of a query is the inverse of the rank of the first hit result.

\paragraph{Probing Tasks}
All the probing tasks we use are classification tasks, and we use classification accuracy as the metric for these tasks.

\subsection{Implementation Details}
Our code is all implemented in Pytorch\footnote{https://pytorch.org/}. We load the pre-trained models from Huggingface\footnote{https://huggingface.co/models} while keeping their hyperparameter settings. Since adapter tuning adjusts fewer parameters than full-model fine-tuning, we set the learning rate of UniXcoder to $5e^{-5}$, and CodeT5 to $1e^{-4}$. We reproduce the results of these models on downstream tasks and present them below.

For the adapter setting, we insert the adapter behind the self-attention layer and feed-forward layer of the encoder and decoder. The dimension of the adapter is set to 128. All experiments are conducted on 4 NVIDIA Tesla V100 cards and each card has 32GB graphic memory.

\section{Experimental Results}
\subsection{RQ1: How Does Multilingual Fine-tuning Perform on Different Models and Downstream Tasks?}
\paragraph{Code Summarization}

\begin{table*}[t]
	\caption{Effectiveness of multilingual fine-tuning for code summarization}
	\begin{center}
		
		\begin{tabular}{l|cccccc|c}
			\hline
			Model&Ruby& JavaScript&Java&Go&PHP&Python&Overall \\
			\hline
			CodeBERT&12.53&  13.86 & 18.72 & 18.15 & 25.48 & 18.25 & 17.83 \\
			$m$CodeBERT& \textbf{14.75}& \textbf{15.80}& \textbf{20.11} & \textbf{18.77} & \textbf{26.23} & \textbf{18.71} & \textbf{19.06}\\
			\hline
			GraphCodeBERT & 12.62&14.79 & 19.22 & 18.40 &  25.45 &  18.02 & 18.08\\
			$m$GraphCodeBERT& \textbf{14.95}&\textbf{15.79} & \textbf{19.91} &  \textbf{18.92} & \textbf{26.15} &  \textbf{18.90} & \textbf{19.10}\\
			\hline
			PLBART &13.94 &  \textbf{16.36} & \textbf{18.73} &  \textbf{17.99} &  \textbf{24.21} &  \textbf{19.79} & \textbf{18.50} \\
			$m$PLBART& \textbf{13.99} & 14.11 & 18.14  & 17.82 &  23.41 & 17.48 & 17.49\\
			\hline
			UniXcoder& \textbf{15.07} &15.69 & \textbf{20.15} & \textbf{19.22} & 26.36 &  19.14 & \textbf{19.27}\\
			$m$UniXcoder& 14.97&\textbf{15.78} & 19.95 &  19.13 & \textbf{26.41} &  \textbf{19.38} &  19.27\\
			\hline
			CodeT5& 15.18 & \textbf{16.09} & \textbf{20.23} & \textbf{19.70} &  \textbf{25.88} &  \textbf{20.26} & \textbf{19.56}\\
			
			$m$CodeT5 & \textbf{15.23} & 15.61& 19.99 & 19.66 &  25.78 & 20.17 & 19.41\\
			\hline
			
		\end{tabular}
		\label{codesum_rq1}
	\end{center}
\end{table*}

\begin{table*}[t]
	\caption{Effectiveness of multilingual fine-tuning for code search}
	\begin{center}
		
		\begin{tabular}{l|cccccc|c}
			\hline
			Model&Ruby& JavaScript&Java&Go&PHP&Python&Overall \\
			\hline
			CodeBERT&67.7&  61.6 & 67.6 &  88.5 &  62.9 &  67.6 & 69.3\\ 
			$m$CodeBERT& \textbf{73.2} & \textbf{64.3} & \textbf{69.7} &  88.5 & \textbf{63.5} & \textbf{67.8} & \textbf{71.2} \\
			\hline
			GraphCodeBERT &70.8 &   64.4 & 69.3 &  89.4 &  \textbf{64.8} &  69.2 & 71.3 \\
			$m$GraphCodeBERT& \textbf{73.8} &   \textbf{66.0} &  \textbf{71.0} &  89.4 & 64.6 & \textbf{69.5} & \textbf{72.4}\\
			
			\hline
			UniXcoder& 73.9 &  \textbf{68.9} & \textbf{72.9} & \textbf{91.6} & \textbf{67.5} & \textbf{72.2} & 74.5\\
			$m$UniXcoder& \textbf{76.4}&68.4 &72.5  &91.2 & 66.8 &  72.0 &  \textbf{74.6}\\
			\hline
			
		\end{tabular}
		\label{codesearch_rq1}
	\end{center}
\end{table*}

In this subsection, we compare the results of multilingual and monolingual fine-tuning on the code summarization task based on different pre-trained models. Monolingual fine-tuning is the original way of fine-tuning one model on the dataset in each language, and multilingual fine-tuning tunes only one model of the same size for all programming languages. The pre-trained models include CodeBERT, GraphCodeBERT, PLBART, UniXcoder, and CodeT5, where CodeT5 is the state-of-the-art model for this task.

The results are shown in Table~\ref{codesum_rq1}. We denote multilingual fine-tuned models with the prefix $m$, as $m$CodeBERT is a multilingual model fine-tuned based on CodeBERT. The results on CodeBERT and GraphCodeBERT are from Ahmed and Devanbu~\cite{Ahmed2022MultilingualTF}. It can be clearly noticed that the results of multilingual fine-tuning based on CodeBERT and GraphCodeBERT are significantly better than monolingual fine-tuning, and multilingual fine-tuning shows its effectiveness in all six programming languages. Overall, the improvements are also significant, with a 6.90\% improvement on CodeBERT and 5.64\% on GraphCodeBERT.

However, on PLBART, UniXcoder, and CodeT5, the results show a different trend. The overall scores of PLBART, UniXcoder, and CodeT5 drop instead. The results of PLBART are from its open source repository\footnote{https://github.com/wasiahmad/PLBART/tree/main/multilingual/}, where the authors conducted exploratory experiments on multilingual code summarization and generation tasks. Results on both tasks show that multilingual fine-tuning leads to performance degradation on most programming languages. On UniXcoder, multilingual tuning causes performance degradation in half of the programming languages. On CodeT5, multilingual tuning only improves on Ruby. 

\paragraph{Code Search}
We also test the effectiveness of multilingual fine-tuning on code search, and the results are shown in Table~\ref{codesearch_rq1}. Since CodeT5 does not report results on code search, we only compare the results based on CodeBERT, GraphCodeBERT, and UniXcoder. UniXcoder is the state-of-the-art model for this task.

As can be seen from Table~\ref{codesearch_rq1}, multilingual fine-tuning shows effectiveness on CodeBERT and GraphCodeBERT, outperforming fine-tuning for each language separately on all programming languages. On UniXcoder, multilingual fine-tuning only improves on Ruby and degrades on other programming languages. Multilingual fine-tuning greatly improves the performance in Ruby, which should be attributed to its data size. Its dataset has the smallest data size, and from Table~\ref{dataset} it can be noticed that datasets of other programming languages are even two to ten times larger than its dataset. 

On both tasks, our experiments reflect similar results. Multilingual fine-tuning is no longer as effective on UniXcoder and CodeT5 as on CodeBERT and GraphCodeBERT. It only shows its superiority in low-resource languages, and this is consistent with the findings of previous studies~\cite{Ha2016TowardMN,Firat2016ZeroResourceTW,Dabre2020ACS} that low-resource languages can benefit through positive knowledge transfer in multilingual learning. While in other programming languages, the results of multilingual fine-tuning are worse than that of monolingual fine-tuning. The reason may be catastrophic forgetting due to learning multiple  programming languages on the same model.

\finding{1}{Based on CodeBERT and GraphCodeBERT, multilingual fine-tuning has shown its effectiveness in all programming languages. On UniXcoder and CodeT5, multilingual fine-tuning benefits low-resource languages and simultaneously results in performance degradation in other programming languages.}

\begin{figure}[t]                         
	\centering                              
	\includegraphics[width=\linewidth]{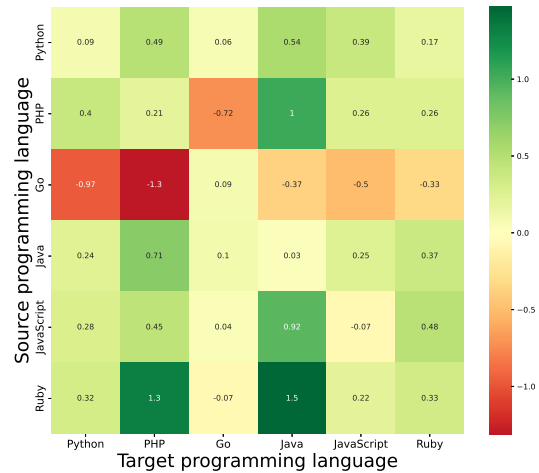} 
	\caption{Relative BLEU-4 improvement of adapter tuning over full-model tuning on cross-lingual code summarization task.}
	\label{corr_codesum}                              
\end{figure} 

\begin{figure}[t]                         
	\centering                              
	\includegraphics[width=\linewidth]{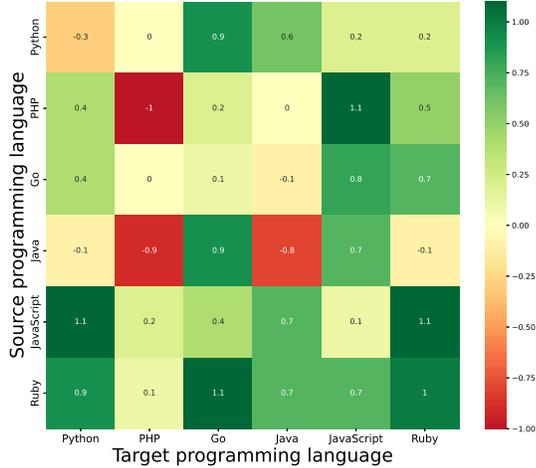} 
	\caption{Relative MRR improvement of adapter tuning over full-model tuning on cross-lingual code search task.}
	\label{corr_codesearch}                              
\end{figure} 

\subsection{RQ2: How Effective is Adapter Tuning in Cross-lingual Scenarios?}

When fine-tuning the entire model for multiple programming languages causes performance degradation due to catastrophic forgetting, we choose adapter tuning to fix most of the pre-trained parameters and update a small number of parameters. We first experiment with applying adapter tuning to cross-lingual scenarios of code summarization and code search.

\paragraph{Code Summarization}
Based on CodeT5, we compare the performance of adapter tuning and full-model fine-tuning on code summarization in cross-lingual scenarios. We fine-tune the model parameters on the dataset in one programming language and evaluate the model on the dataset in the other. To get a visual impression of the performance difference, Fig.~\ref{corr_codesum} shows the relative BLEU-4 improvement of adapter tuning over full-model fine-tuning. The vertical axis is the programming language of the training set, and the horizontal axis is the programming language for evaluation. Fig.~\ref{corr_codesum} shows that adapter tuning performs better than full-model fine-tuning on most cross-lingual tasks. The exception is that adapter tuned in Go language performs worse in most programming languages than full-model fine-tuning. It can be noted that adapter tuning in other programming languages does not benefit significantly in Go language either. For adapter, data in Go language are not easy to adapt. More data or parameters may need to be involved.

\begin{table*}[t]
	\caption{Code Summarization Comparison}
	\begin{center}
		\begin{tabular}{lc|cccccc|c}
			\hline
			Model&Parameters&Ruby&JavaScript&Java&Go&PHP&Python&Overall\\
			\hline
			CodeBERT& 6x 173M & 12.16 & 14.90 & 17.65 &18.07 & 25.16 & 19.06& 17.83 \\
			GraphCodeBERT & 6x 173M & 12.39 & 14.81 & 19.00 & 18.41 & 25.59 & 18.06 & 18.04 \\
			PLBART & 6x 140M & 14.11 & 15.56 & 18.45 & 18.91 & 23.58 & 19.30 & 18.32 \\
			\hline
			UniXcoder&6x 253M& 15.07&15.69& \textbf{20.15}& 19.22 & 26.36&19.14&19.27 \\
			$m$Adapter-UniXcoder & 10M& \textbf{15.43} &\textbf{15.87} & 20.01 & \textbf{19.49} & \textbf{26.46} & \textbf{19.83} & \textbf{19.52}\\
			$m$UniXcoder& 263M& 14.97&15.78 & 19.95 &  19.13 & 26.41 &  19.38 &  19.27\\
			\hdashline
			\multicolumn{2}{c|}{p-value}& \textless 0.001 & \textless 0.001 & 0.440 & \textless 0.001 & 0.386 & \textless 0.001 & \textless 0.001 \\
			\hline
			CodeT5 & 6x 223M &  15.18 & \textbf{16.09} & 20.23 & 19.70 &  25.88 &  20.26 & 19.56\\
			$m$Adapter-CodeT5 &9M&   \textbf{15.49} & 16.06 &  \textbf{20.42} &  \textbf{19.84} &  \textbf{26.08} &  \textbf{20.52} &  \textbf{19.74}\\
			$m$CodeT5&232M & 15.23&15.61&19.99&19.66&25.78&20.17&19.41 \\
			\hdashline
			\multicolumn{2}{c|}{p-value}& \textless 0.001 & \textless 0.001 & \textless 0.001 & \textless 0.001 & \textless 0.001 & \textless 0.001 & \textless 0.001 \\
			\hline
			
		\end{tabular}
		\label{codesum_rq3}
	\end{center}
\end{table*}

\begin{table*}[t]
	\caption{Code Search Comparison}
	\begin{center}
		\begin{tabular}{lc|cccccc|c}
			\hline
			Model&Parameters&Ruby&JavaScript&Java&Go&PHP&Python&Overall\\
			\hline
			CodeBERT& 6x 125M & 67.9 & 62.0 &67.6 & 88.2 & 62.8 & 67.2 & 69.3\\
			GraphCodeBERT & 6x 125M & 70.3 & 64.4 & 69.1 & 89.7 & 64.9 & 69.2 & 71.3 \\
			PLBART & 6x 140M & 67.5 & 61.6 & 66.3 & 88.7 & 61.1 & 66.3 & 68.5 \\
			\hline
			UniXcoder  & 6x 126M& 73.9 & 68.9 & 72.9 & \textbf{91.6} & \textbf{67.5} & 72.2&74.5 \\
			$m$Adapter-UniXcoder & 5M & \textbf{77.3} & \textbf{70.2} & \textbf{73.5} & 90.9 & 67.1 & \textbf{72.7}&\textbf{75.3} \\
			$m$UniXcoder & 131M & 76.4 & 68.4 & 72.5 & 91.2 & 66.8 & 72.0 & 74.6 \\
			\hdashline
			\multicolumn{2}{c|}{p-value}&0.004 & \textless 0.001 & \textless 0.001 &0.986 &0.445 & \textless 0.001 & \textless 0.001\\
			\hline
			
		\end{tabular}
		\label{codesearch_rq3}
	\end{center}
\end{table*}
\paragraph{Code Search}
On code search task, we also test the relative performance of adapter tuning versus full-model fine-tuning based on UniXcoder, as shown in Fig.~\ref{corr_codesearch}. In most cross-lingual tasks, adapter tuning outperforms full-model fine-tuning with fewer parameter updating. The tasks where adapter tuning performs worse than full-model fine-tuning are mainly distributed in the diagonal part of the figure. These tasks are trained and evaluated in the same language and are not cross-lingual tasks. On these monolingual tasks, full-model fine-tuning allows more parameters to be adjusted to fit the task than adapter tuning, without the concern of catastrophic forgetting.

\finding{2}{Adapter tuning is more effective than full-model fine-tuning in most cross-lingual scenarios on code search and summarization tasks.}

\begin{table*}[t]
	\caption{Results of multilingual tuning in low-resource code summarization task}
	\begin{center}
		\begin{tabular}{c|cccccc|c}
			\hline
			Training samples&Ruby&JavaScript&Java&Go&PHP&Python&Overall\\
			\hline
			6x 100& 12.81 & 13.41 & 12.44 & 14.89 & 19.35 & 15.28 & 14.70\\
			6x 200 & 15.03 & 15.24 & 18.92 & 16.38 & 23.65 & 18.85 & 18.01 \\
			6x 500 & 15.21 & 15.64 & 19.28 & 18.34 & 24.76 & 19.02 & 18.71 \\
			6x 1,000 & 15.32 & 15.62 & 19.36 & 18.79 & 24.72 & 19.26 & 18.85\\
			\hline
			908,224&  \textbf{15.49} & \textbf{16.06} &  \textbf{20.42} &  \textbf{19.84} &  \textbf{26.08} &  \textbf{20.52} &  \textbf{19.74}\\
			\hline
			
		\end{tabular}
		\label{rq4}
	\end{center}
\end{table*}

\subsection{RQ3: How Effective is Adapter Tuning Over Multilingual Full-model Fine-tuning on Different Pre-trained Models?}

When adapter tuning demonstrates its effectiveness in cross-lingual scenarios, we further explore its performance in multilingual tasks. 

\paragraph{Code Summarization}

Table~\ref{codesum_rq3} shows the comparison results on the code summarization task. The model with prefix $m$ is a multilingual model that requires only one set of parameters and the other models have to train models for six programming languages separately. Based on UniXcoder and CodeT5, adapter tuning shows its effectiveness in all programming languages compared to multilingual full-model fine-tuning. Compared to original fine-tuning, $m$adapter also outperforms various monolingual models in most programming languages with fewer parameter updating.

In order to verify whether the improvement of $m$adapter on pre-trained models over multilingual full-model fine-tuning is significant, we apply one-sided pairwise t-test for statistical analysis. The null hypothesis is rejected for all six languages on CodeT5 and most languages on UniXcoder. It is evident that $m$adapter has a statistically significant improvement over multilingual training on code summarization.

\paragraph{Code Search}
On code search, as shown in Table~\ref{codesearch_rq3}, adapter tuning also outperforms monolingual and multilingual full-model fine-tuning in most  programming languages. The improvement is particularly significant for low-resource programming languages, such as Ruby and JavaScript. Statistical analysis results show that the improvement of $m$adapter over multilingual training is statistically significant (p \textless 0.001) for all programming languages except Ruby with a 0.004 p-value.

\finding{3}{Although fewer parameters are updated, adapter tuning is more effective in multilingual learning than full-model fine-tuning. Moreover, it also outperforms the results of fine-tuning separately for each programming language.}

\subsection{RQ4: How Effective is Multilingual Fine-tuning in Low-resource Scenarios?}

Multilingual models are fine-tuned from data of multiple programming languages. Joint training and consequent positive transfer benefit the learning of various programming languages. In reality, many programming languages lack high-quality labeled data. Therefore, we evaluate whether a well-performing multilingual model can be learned with extremely limited data for each programming language.

We sample training examples equally from datasets of six programming languages, and gather multiple multilingual datasets of sizes 600, 1200, 3000 and 6000. We fine-tune $m$Adapter-CodeT5 on these datasets and compare the results with adapter tuning on the whole dataset. The results are shown in Table~\ref{rq4}. 

The table shows that as the number of training samples increases, the performance of adapter tuning continues to improve. There is a significant difference between training results using 100 samples for each programming language and the other results. At this point, the model fails to converge due to a lack of training data. When training with 200 samples per programming language, there is a difference of fewer than 2 BLEU-4 from fine-tuning the entire dataset. When the number of samples is increased to 1000 for each language, the difference with the baseline is less than 1 BLEU-4. It is evident that multilingual training is effective for the low-resource code summarization task.

It can be noticed that the growth of model performance gradually slows down as the data increase. In the comparison between the last row and the penultimate row, even more than 900,000 samples bring an improvement of no more than 1 BLEU-4 score. This shows, on the one hand, that the pre-trained model is robust enough and can be quickly adapted to the downstream task with very few data. On the other hand, it demonstrates the potential of multilingual fine-tuning, which can make full use of multilingual data for rapid convergence of the model in low-resource scenarios. 

\finding{4}{Multilingual fine-tuning is so effective in the low-resource code summarization task that it is possible to approach the results of fine-tuning with the entire dataset using very few samples per programming language.}

\subsection{RQ5: Why is Adapter Tuning Better than Full-model Fine-tuning in the Above Scenarios?}

Adapter tuning demonstrates its effectiveness in cross-lingual, multilingual and low-resource scenarios by updating only few parameters. To inspect models at a fine-grained level and check whether the adapter behaves consistently, we apply probing experiments to examine whether models encode a set of code characteristics. We adopt LEN, CPX and TYP tasks to probe for code surface, structural and semantic information, respectively.

Specifically, we insert the adapter to BERT~\cite{Devlin2019BERTPO}, CodeBERT, GraphCodeBERT and UniXcoder, and fine-tune them on code search along with the original models. Then we extract feature vectors from the hidden layers of these models and train a simple linear classifier to associate these feature vectors with different code characteristics. Since the linear classifier has no hidden units, its performance on probing tasks heavily depends on the feature vectors from these models.

\begin{figure*}[t]                         
	\centering                              
	\includegraphics[width=\linewidth]{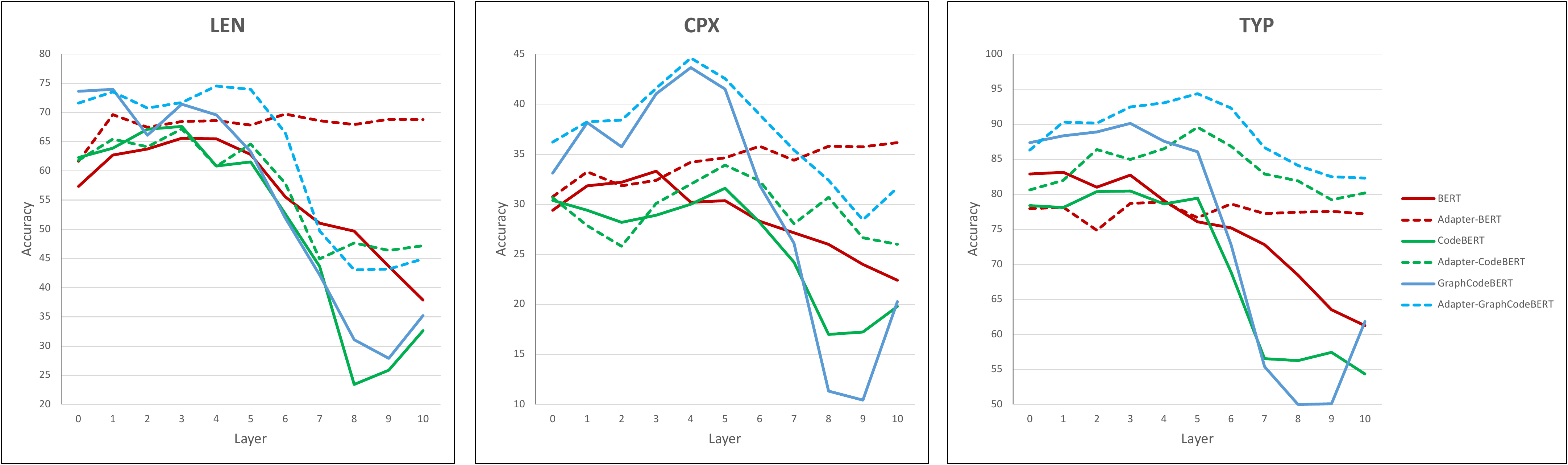} 
	\caption{Accuracy of different models on LEN, CPX and TYP tasks. The horizontal axis indicates the index of the hidden layer used for probing.}
	\label{probe_fig}                              
\end{figure*} 
\begin{table}[t]
	\caption{Probing Task accuracy}
	\begin{center}
		\begin{tabular}{l|ccc}
			\hline
			\multirow{2}{*}{Model}&LEN& CPX &TYP\\
			&surface& structural &semantic\\
			\hline
			BERT& 23.85 &  15.55 & 54.10\\
			Adapter-BERT& \textbf{65.20} & 37.00 & 79.70\\
			\hline
			CodeBERT&  24.15 & 17.70 & 53.60\\
			Adapter-CodeBERT&  57.95& 35.65 & 87.50\\
			\hline
			GraphCodeBERT& 35.25 & 24.30 & 61.85\\
			Adapter-GraphCodeBERT&44.95& \textbf{40.35} & 82.30\\
			\hline
			UniXcoder& 47.25 & 31.30 & 86.70\\
			Adapter-UniXcoder& 52.55 & 36.25 & \textbf{92.65}\\
			\hline
			
		\end{tabular}
		\label{probing_result}
	\end{center}
\end{table}

We extract feature vectors from the final hidden layers of these models for probing, and the results are presented in Table~\ref{probing_result}. Overall, these models perform best on invalid type prediction. Adapter-UniXcoder achieves 92.65\% accuracy on this task, indicating that the model does encode such semantic information. Adapter-BERT performs the best on code length prediction with 65.20\% accuracy. This task essentially predicts the number of tokens in the input sequence, and the additional pre-training task of other models may hurt this task. The finding is also consistent with the conclusion found by Karmakar and Robbes~\cite{Karmakar2021WhatDP} in their probing experiments. On the cyclomatic complexity task, Adapter-GraphCodeBERT is the best-performing model. Since GraphCodeBERT is specifically pre-trained on the structural information, this result is not unexpected. However, its advantage over other models is not obvious, perhaps because the task itself is more challenging.

Further comparing models fine-tuned with the adapter to the original models in Table~\ref{probing_result}, there is a pretty significant difference in their performance, especially on BERT, CodeBERT and GraphCodeBERT. On all probing tasks, models with adapter tuning show a clear superiority over the original models.

To more clearly compare adapter tuning with their original models, we also extract hidden representations of all previous layers as feature vectors and evaluate their performance on probing tasks. Fig.~\ref{probe_fig} shows how the performance on three probing tasks varies with the hidden layer index on BERT, CodeBERT and GraphCodeBERT.\footnote{The detailed results on all models are shown in the following anonymous repository.} 

The dashed lines in the figure represent the performance of models with adapter tuning in probing tasks. The solid lines of the same colour represent the performance of the corresponding original models. Overall, although each probing task corresponds to different code characteristics for examination, there is a great deal of similarity in the accuracy variation across models on these tasks. In the first six hidden layers, there is little difference between the performance of adapter tuning and full-model fine-tuning. In the subsequent hidden layers, the performance gap between adapter tuning and original fine-tuning becomes increasingly prominent.

\begin{table*}[t]
	\caption{Comparison of different adapter designs on code summarization and code search}
	\begin{center}
		\begin{tabular}{l|c|cccccc|c}
			\hline
			Task&Model&Ruby&JavaScript&Java&Go&PHP&Python&Overall\\
			\hline
			\multirow{2}{*}{Code Summarization} & $m$Adapter-CodeT5 &  15.49 &  \textbf{16.06} &  \textbf{20.42} &  \textbf{19.84} &  \textbf{26.08} &  \textbf{20.52} &  \textbf{19.74}\\  
			&$m$Adapter-MoE-CodeT5 &  \textbf{15.62} & 16.04 & 20.03 & 19.79 & 25.75 & 20.36 & 19.60\\
			\hline
			\multirow{2}{*}{Code Search}&$m$Adapter-UniXcoder  & \textbf{77.3} & \textbf{70.2} & \textbf{73.5} & \textbf{90.9} & \textbf{67.1} & \textbf{72.7}&\textbf{75.3} \\
			& $m$Adapter-MoE-UniXcoder &76.6 & 69.1 & 72.8 & 90.7 & 66.3 & 72.1 & 74.6 \\
			\hline
			
		\end{tabular}
		\label{adapter_design}
	\end{center}
\end{table*}

\begin{table*}[t]
	\caption{Impact of different mini-batches on code summarization and code search}
	\begin{center}
		\begin{tabular}{l|cl|cccccc|c}
			\hline
			Task&\multicolumn{2}{c|}{Method}&Ruby&JavaScript&Java&Go&PHP&Python&Overall\\
			\hline
			\multirow{2}{*}{Code Summarization} & \multirow{2}{*}{$m$Adapter-CodeT5}& - Multilingual &  15.49 &  \textbf{16.06} &  \textbf{20.42} &  \textbf{19.84} &  \textbf{26.08} &  \textbf{20.52} &  \textbf{19.74} \\
			& & - Monolingual& \textbf{15.66} & 16.03&20.14&19.76&25.63& 20.29& 19.59\\
			\hline
			\multirow{3}{*}{Code Search}&\multirow{3}{*}{$m$Adapter-UniXcoder} & - Multilingual &  76.7& 69.3&73.1&90.7&66.5&72.2&74.8\\
			& & - Monolingual& \textbf{77.3} & \textbf{70.2} & \textbf{73.5} & \textbf{90.9} & \textbf{67.1} & \textbf{72.7}&\textbf{75.3}\\
			& & - w/o tags & 76.6 & 69.9 & 73.1 & 90.7 & 66.9 & 72.6&75.0\\
			\hline
			
		\end{tabular}
		\label{mini-batch_design}
	\end{center}
\end{table*}

It should be noted that these tasks are fine-tuned for code search. However, code characteristics needed for these probing tasks are not always helpful for the downstream task. Therefore, the accuracy of different models on probing tasks decreases on the last hidden layers. Since adapter tuning performs significantly better than full-model fine-tuning in the last hidden layers, models with adapter tuning encode more information than origin models. In contrast, full-model fine-tuning suffers from catastrophic forgetting and discard this information. These code characteristics are low-level information and generalized across languages than that required for downstream tasks. Therefore, adapter tuning can perform better on cross-lingual and multilingual tasks with the help of this information. We conjecture that this is why adapter tuning can mitigate catastrophic forgetting issues.

From overall changes of the accuracy, it can be noticed that the accuracy on LEN decreases with the depth of hidden layers. In contrast,  the accuracy of CPX increases first and then decreases gradually. Except for natural language pre-trained model BERT, the best performance of CodeBERT and GraphCodeBERT is also achieved in the middle hidden layer on TYP. Since LEN corresponds to probing surface information and CPX and TYP correspond to structural and semantic information of the code, it is clear that these models learn surface information in the lower layers and structural and semantic features in the deeper layers. This conclusion is consistent with previous studies.

From the perspective of each task, Adapter-BERT consistently maintains high accuracy on LEN, while all other models discard some of the surface information. This probing task shows the difference between source code and natural language pre-trained models in handling code downstream tasks. On the CPX task, GraphCodeBERT maintains high accuracy in the shallow hidden layers, which confirms that GraphCodeBERT adequately encodes that structural information. However, the performance of CPX drops significantly in the last hidden layers. The downstream task may not require the involvement of cyclomatic complexity information. On TYP task, different models with adapter tuning end up with high accuracy, and that semantic information may be effectively involved in the downstream task. The variation in accuracy is consistently best on GraphCodeBERT, followed by CodeBERT and then BERT. This also coincides with the performance of pre-trained models in the downstream task.

\finding{5}{Adapter tuning significantly outperforms full-model fine-tuning on all probing tasks. In addition, we observe that the hidden layers of these models gradually encode higher-level information from shallow to deep, consistent with previous studies.}

\section{Analysis}
In this section, we consider several factors that significantly impact the results, including the adapter design, data batch, and the adapter dimension.

\subsection{One or Multiple Adapters?}

This study employs a multilingual model to handle tasks in multiple programming languages.  Therefore, mutual interference within the model is inevitable when dealing with tasks in different programming languages. To alleviate this issue, we take inspiration from the Sparsely-Gated Mixture-of-Expert (MoE) layer~\cite{Shazeer2017OutrageouslyLN} and split the original 128-dimensional adapter into four 32-dimensional adapters. Without significantly increasing the parameters, we expect the model to learn multiple adapters as experts dealing with data in different programming languages separately. We use a gating network to implement the selection of adapters and select the two best-performing adapters to participate in the computation when processing different samples.

Based on two pre-trained models, UniXcoder and CodeT5, we compare the performance of this implementation with that of the original adapter on code search and code summarization. The experimental results are shown in Table~\ref{adapter_design}. The method performs worse than the original adapter on both tasks. We conjecture that there is not enough information to guide the model to select different adapters as experts, and more data may be needed to support this.

\begin{table*}[t]
	\caption{Impact of adapter bottleneck dimension size on code summarization and code search}
	\begin{center}
		\begin{tabular}{l|cc|cccccc|c}
			\hline
			Task&Method &Dimension&Ruby&JavaScript&Java&Go&PHP&Python&Overall\\
			\hline
			\multirow{3}{*}{Code Summarization} & \multirow{3}{*}{$m$Adapter-CodeT5} & 24 & \textbf{15.82} & 15.95 & 18.76 & 15.32 & 24.34 & 20.35 & 18.42\\
			& & 64 &15.69 & 16.03 & 20.42 & 19.73 & 25.54 & 20.02 & 19.57\\
			&&128 &  15.49 &  \textbf{16.06} &  \textbf{20.42} &  \textbf{19.84} &  \textbf{26.08} &  \textbf{20.52} &  \textbf{19.74} \\

			\hline
			\multirow{3}{*}{Code Search}&\multirow{3}{*}{$m$Adapter-UniXcoder} & 24 &  73.9 & 66.5 & 70.7 & 87.9 & 64.7 & 70.1 & 72.3\\
			& & 64 & 76.8 & 69.6 & 73.0 & 90.7 & 66.5 & 72.3& 74.8\\
			& & 128 & \textbf{77.3} & \textbf{70.2} & \textbf{73.5} & \textbf{90.9} & \textbf{67.1} & \textbf{72.7}&\textbf{75.3}\\
			\hline
			
		\end{tabular}
		\label{dimension_design}
	\end{center}
\end{table*}

\subsection{Monolingual Mini-batches vs Multilingual Mini-batches}

When learning multilingual data, an interesting question is whether it is better for the model to learn samples of the same language each time or to learn random samples. Specifically, when optimizing a model using stochastic gradient descent, gradients are computed over mini-batches. The samples of a mini-batch are all the data that the model is exposed to in one training step. A better mini-batch choice would facilitate model convergence, which in turn improves performance on the downstream task.

We test two mini-batch settings. One uses multilingual mini-batch, which randomly sorts multilingual datasets and then divides mini-batches. The other applies monolingual mini-batches by dividing mini-batches on each monolingual dataset and then randomly ordering these mini-batches. We evaluate these settings on both tasks and models. The experimental results are shown in Table~\ref{mini-batch_design}. 

On code summarization, CodeT5 trained with multilingual mini-batches outperforms the same model with monolingual mini-batches. On code search, the experimental finding of UniXcoder is the opposite. However, Ahmed and Devanbu~\cite{Ahmed2022MultilingualTF} find that the performance of multilingual mini-batch setting is always better in multilingual full-model fine-tuning. We argue that the phenomenon is task-related. The code search task may require more programming language category information to match code and queries. In contrast, the goal of multilingual code summarization task is about generating natural language descriptions. We further introduce language-specific tags as prompts into code search task to supplement the programming language category information. Experimentally, we find that adding language-specific tags does improve the performance.

\subsection{The Dimension of Adapters}

An essential parameter of the adapter is its bottleneck dim, which determines the capacity of the adapter. In this paper, we vary the adapter bottleneck dim from 24 to 128 and conduct experiments on all tasks and models. Table~\ref{dimension_design} shows that increasing the bottleneck dimension of adapters can significantly improve the performance of most tasks on UniXcoder and CodeT5. The model performs better in Ruby for code summarization on smaller adapter dimensions. We speculate that this low-resource language mainly benefits from the data in other programming languages and also gives way to learning  other programming languages due to insufficient data. A more balanced learning approach for multilingual data may be required. In this study, we set the adapter size to 128. A larger adapter capacity may bring more performance improvement, but we leave it for future work because of limited computing resources.

\section{Related Work}

\subsection{Pre-training for Programming Language}
With the great success of pre-trained models in the field of natural language processing, a range of pre-trained models in programming languages have arisen to facilitate source code understanding and generation tasks, such as code search, code generation, and bug detection~\cite{Wang2022BridgingPM,Liu2020BugSumDC,Wan2018ImprovingAS,Wang2020ModularTN,Gu2018DeepCS,Lam2017BugLW}. CuBERT~\cite{Kanade2020PretrainedCE} and CodeBERT are first proposed to learn representations of programming languages using large-scale unlabeled data in a self-supervised way. CuBERT utilizes the mask language modeling pre-training objective in BERT, and CodeBERT is pre-trained in both natural and programming languages. GraphCodeBERT  introduces data flow information on top of CodeBERT to facilitate the understanding of code structure.  Apart from the aforementioned encoder-only models, decoder-only models are also proposed for programming languages. GPT-C~\cite{Svyatkovskiy2020IntelliCodeCC} and CodeGPT~\cite{Lu2021CodeXGLUEAM} both utilize unidirectional language modeling that uses all previous tokens to predict the next token for pre-training.

Some recent works explore encoder-decoder models to support both understanding and generation tasks, including PLBART, CodeT5 and UniXcoder. PLBART is based on BART  and pre-trained with denoising objectives. CodeT5 adapts T5~\cite{Raffel2020ExploringTL} to solve code-related tasks and allows for multi-task learning for various downstream tasks. UniXcoder is based on UniLM~\cite{Dong2019UnifiedLM} and pre-trained on multi-modal data, including code, comment, and AST. In this paper, we conduct experiments on the recent models UniXcoder and CodeT5.

\subsection{Adapter Tuning}
Adapter modules are initially investigated in computer vision tasks and have been used to adapt models for multiple domains~\cite{Rebuffi2017LearningMV}. In NLP, adapters are used for parameter efficient fine-tuning of the base pre-trained Transformer model to adapt to new tasks~\cite{Stickland2019BERTAP,Houlsby2019ParameterEfficientTL}. Santoro~\etal~apply adapter modules to avoid catastrophic forgetting~\cite{Santoro2016OneshotLW}. Bapna and Firat use adapters to fine-tune a multilingual translation model in various languages~\cite{Platanios2018ContextualPG}. MAD-X pre-trains task-specific and language-specific adapters and then combines their representations to exploit cross-task and linguistic knowledge~\cite{Pfeiffer2020MADXAA}. There are also research works showing that adapter-based fine-tuning can perform better than normal fine-tuning on few-shot and cross-lingual scenarios~\cite{He2021OnTE} and is more robust under adversarial attacks~\cite{Han2021RobustTL}. In this paper, we try to adapt adapters for source code understanding and generation tasks in multiple programming languages.

\section{Threats to Validity}

\textbf{External Validity.} In this study, we experiment on a limited number of tasks, datasets, and pre-trained models. These may all bring bias to the results. We select representative and state-of-the-art models and the most widely-used datasets to mitigate this issue. In the open source repository, we complement the experiments of adapter tuning with CodeBERT and GraphCodeBERT on code summarization and code search. The results are still promising and consistent with UniXcoder and CodeT5. We also maintain the same random seeds across models to ensure the consistency of the experiments. In the future, we would extend adapter tuning to more code intelligence tasks and datasets. Also,  probing experiments are targeted at several local aspects of the code information, and the final conclusion is drawn by induction rather than strict proof.

\textbf{Internal Validity.} 
We illustrate the effectiveness of adapter tuning in various scenarios. However, the adapter applied in our paper is the standard structure. In fact, we try to modify the structure of the adapter or to extend the capacity of adapter using mixture-of-experts, but none of these attempts achieve the desired result. Adapter can achieve better results with few parameter tuning, which should have more exciting findings to explore.

\section{Conclusion}
We start this study with performance degradation of multilingual fine-tuning and explore the performance of adapter tuning in cross-lingual, multilingual  and low-resource scenarios. We experimentally demonstrate the effectiveness of adapter tuning in these scenarios. It outperforms full-model fine-tuning with fewer parameter updating in all scenarios. We find that multilingual fine-tuning with 200 random samples per programming language can approach the performance of training on the entire dataset. When multilingual full-model fine-tuning suffers from catastrophic forgetting, we demonstrate through probing experiments that adapter tuning can overcome this issue significantly.

As pre-trained models become increasingly huge, the adapter as a small neural module provides a way to simplify and accelerate transfer learning. In the future, we would like to explore the integration of code knowledge into the adapter and the performance of the adapter in more code intelligence tasks. Our source code and models are publicly available at:~\url{https://github.com/wangdeze18/Multilingual-Adapter-for-SE}.

\section*{Acknowledgment}
The authors would like to thank the anonymous reviewers for their insightful comments. This work was substantially supported by National Natural Science Foundation of China (No. 62032019, 61872373, and 62272473). This work was also supported by National Key R\&D Program of China 2022YFC3400404.

\bibliographystyle{IEEEtran}
\bibliography{IEEEabrv,IEEEexample}

\end{document}